\documentclass[letterpaper,journal]{IEEEtran}
\usepackage[left=1.7cm,right=1.7cm,top=1.7cm, bottom=1.7cm]{geometry}
\usepackage{cite}
\usepackage{amsmath,amssymb,amsfonts,bm,mathtools}
\usepackage{graphicx}
\usepackage{balance}
\usepackage{color}
\usepackage{url}
\usepackage{hyperref}
\usepackage{comment}
\usepackage{subcaption}
\usepackage{multirow}
\usepackage{siunitx}

\makeatletter
\@ifundefined{lemma}{}{}
\@ifundefined{theorem}{}{}
\@ifundefined{proposition}{\newtheorem{proposition}{Proposition}}{}
\@ifundefined{corollary}{}{}
\makeatother
\newcommand{\CN}{\mathcal{CN}}
\newcommand{\E}{\mathbb{E}}
\newcommand{\Cov}{\mathrm{Cov}}

\newcommand{\vect}{\mathrm{vec}}
\newcommand{\tr}{\mathrm{tr}}

\newcommand{\Real}{\mathrm{Re}}
\newcommand{\Imag}{\mathrm{Im}}
\newcommand{\Ind}{\mathcal{X}}
\newcommand{\IndTest}{\mathcal{X}_\ast}

\newcommand{\orcidauthorA}{\href{https://orcid.org/0000-0001-5792-0842}{\includegraphics[scale=0.05]{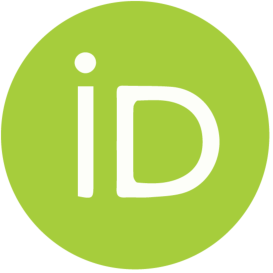}}}
\newcommand{\orcidauthorB}{\href{https://orcid.org/0000-0002-1478-2272}{\includegraphics[scale=0.05]{Part_II_Figures/Shah_WCL_fig_orcid.eps}}}
\newcommand{\orcidauthorC}{\href{https://orcid.org/0000-0003-4363-396X}{\includegraphics[scale=0.05]{Part_II_Figures/Shah_WCL_fig_orcid.eps}}}

\usepackage[font=small]{caption}

\begin{document}

\title{Improved GPR-Based CSI Acquisition via Spatial-Correlation Kernel}
\author{{Syed Luqman Shah$^{\orcidauthorA{}}$, \textit{Graduate Student Member}, \textit{IEEE},  Nurul Huda Mahmood$^{\orcidauthorB{}}$, \textit{Member}, \textit{IEEE}, \\ and Italo Atzeni$^{\orcidauthorC{}}$, \textit{Senior Member}, \textit{IEEE}}\\
    \thanks{The authors are with Centre for Wireless Communications, University of Oulu, Finland (e-mail: \{syed.luqman, nurulhuda.mahmood, italo.atzeni\}@oulu.fi.) This work was supported by the Research Council of Finland (336449 Profi6, 348396 HIGH-6G, 359850 6G-ConCoRSe, and 369116 \href{https://www.6gflagship.com/}{6G Flagship}). Reproducible code for this work available at:~\url{https://github.com/Syed-Luqman-Shah-19/MIMOGPR2}.}}

\maketitle
\begin{abstract}
Accurate channel estimation with low pilot overhead and computational complexity is key to efficiently utilizing multi-antenna wireless systems. Motivated by the evolution from purely statistical descriptions toward physics- and geometry-aware propagation models, this work focuses on incorporating channel information into a Gaussian process regression (GPR) framework for improving the channel estimation accuracy. In this work, we propose a GPR-based channel estimation framework along with a novel \textit{Spatial-Correlation} (SC) kernel that explicitly captures the channel's second-order statistics. We derive a closed-form expression of the proposed SC-based GPR estimator and prove that its posterior mean is optimal in terms of linear minimum mean-square error (LMMSE) under the same second-order statistics, without requiring the underlying channel distribution to be Gaussian. Our analysis reveals that, even with a $50\%$ pilot overhead reduction, the proposed method achieves the lowest normalized mean-square error, competitive empirical $95\%$ credible-interval coverage, and superior preservation of spectral efficiency compared to benchmark estimators, while maintaining lower computational complexity than the conventional LMMSE estimator.
\end{abstract}


\section{Introduction}
\IEEEPARstart{G}{aussian} process regression (GPR) has emerged as a promising approach for channel state information (CSI) acquisition under tight pilot budgets, a critical requirement for 6G-and-beyond wireless networks~\cite{shah2025LOWoverhead}. Conventional least-squares (LS) and linear minimum mean-square error (LMMSE) estimators require pilot lengths that scale with the number of transmit antennas, leading to substantial overhead for large arrays~\cite{Jieao2025IEEE_TIT, ShahNovel}. The LMMSE estimator is computationally expensive for such setups, which has motivated the development of lower-complexity alternatives~\cite{Li2025HolographicLEO}. Compressed-sensing-based methods alleviate overhead by exploiting angular or delay-domain sparsity~\cite{Matrix_Completion_Zhiwei2025, ShahNovel}, but they remain sensitive to noise and low-rank model mismatch in rich-scattering environments~\cite{ADMM_algo_Sig_letter_2018}. These limitations underscore the need for GPR-based CSI acquisition, which offers improved estimation accuracy with reduced pilot overhead~\cite{Jieao2025IEEE_TIT, shah2025LOWoverhead, ShahNovel}.


GPR is a non-parametric Bayesian framework for channel estimation that encodes prior knowledge through a covariance function (kernel). Its posterior mean yields the channel estimate, while the posterior covariance quantifies uncertainty~\cite{GPs_For_ML, ShahNovel}. GPR has been applied to interference prediction and proactive resource allocation~\cite{shah2025}, hardware distortion modeling~\cite{Arj26}, and CSI forecasting~\cite{li2024spatioGPR}. More recently,~\cite{ShahNovel, shah2025LOWoverhead} modeled wireless propagation over a two-dimensional antenna grid and showed that observing only a subset of transmit-antenna responses can reconstruct unobserved channel entries. Despite these advances, existing GPR-based channel estimation methods~\cite{Jieao2025IEEE_TIT,ShahNovel,shah2025LOWoverhead} still have three key limitations: i) marginal-likelihood hyperparameter learning scales poorly with antenna dimension~\cite{GPs_For_ML}; ii) distance-based kernels often assume stationarity and weak anisotropy, causing model mismatch under direction-dependent scattering and non-separable transmit-receive coupling~\cite{demir2024spatial}; and iii) prior second-order channel statistics are not explicitly exploited under sparse training, limiting the use of known covariance structures~\cite{shah2025LOWoverhead}.


To overcome these limitations, this work extends the reduced-pilot GPR framework in~\cite{shah2025LOWoverhead} through a novel \textit{Spatial-Correlation} (SC) kernel induced directly by the channel covariance. In contrast to generic distance-based kernels, the proposed SC kernel is grounded in established channel models, thereby removing marginal-likelihood hyperparameter learning and enabling markedly more efficient and accurate inference under sparse training. We prove that the proposed SC kernel is positive semidefinite (PSD) and derive a closed-form SC-GPR estimator whose posterior mean is LMMSE-optimal under the same second-order statistics. 

In addition to achieving the LMMSE mean, the proposed SC-GPR estimator offers the following advantages: i) reduced pilot overhead; ii) uncertainty-aware channel estimates via credible intervals; and iii) lower computational cost. The proposed SC-GPR framework therefore provides a unified and physically grounded solution for covariance-aware reduced-pilot CSI acquisition, uncertainty-aware reconstruction via credible intervals, and scalable implementation. We validate the proposed method using two closed-form channel covariance families under three reduced-overhead schemes corresponding to $50\%$, $67\%$, and $75\%$ pilot reduction. Numerical results show that SC-GPR achieves lower normalized mean-square error (NMSE) than the benchmark schemes even at $50\%$ pilot reduction, and further delivers superior spectral efficiency (SE) under $75\%$ pilot reduction while maintaining lower computational complexity than the optimal LMMSE estimator.


\section{System Model}
We consider a narrowband point-to-point MIMO link with $N_{\textrm{t}}$ transmit antennas and $N_{\textrm{r}}$ receive antennas. During the channel estimation phase, the pilot overhead is reduced by activating only a subset of the $N_{\textrm{t}}$ antennas instead of sounding the entire array. Specifically, only $n_{\textrm{t}} < N_{\textrm{t}}$ transmit antennas are activated, which relaxes the requirement on the pilot length from $T \geq N_{\textrm{t}}$ to $T \geq n_{\textrm{t}}$~\cite{shah2025LOWoverhead}. The pilots transmitted by these $n_{\textrm{t}}$ active antennas are received across all the $N_{\textrm{r}}$ receive antennas~\cite{ShahNovel}.
Let $\Omega_{\textrm{t}}=\{a_1,\ldots,a_{n_{\textrm{t}}}\}\subset\{1,\ldots,N_{\textrm{t}}\}$ denote an ordered index set corresponding to the $n_{\textrm{t}}$ antennas, with no repeated elements. We adopt a structured equispaced activation pattern with stride $\Delta \in \mathbb{N}$, with
$a_i = 1 + (i-1)\Delta$ for $i=1,\ldots,n_{\textrm{t}}$, where $n_{\textrm{t}}=\left\lfloor (N_{\textrm{t}}-1)/\Delta\right\rfloor+1$ is the maximum integer satisfying the array-size constraint
$1+(n_{\textrm{t}}-1)\Delta \le N_{\textrm{t}}$.

Let $\mathbf{S}\in\mathbb{C}^{n_{\textrm{t}}\times T}$ denote the pilot matrix transmitted by the $n_{\textrm{t}}$ transmit antennas, whose rows are orthonormal and satisfy $\mathbf{S}\mathbf{S}^{\mathsf H}=\mathbf{I}_{n_{\textrm{t}}}$. Based on $\Omega_{\textrm{t}}$, we define the antenna-selection matrix $\mathbf{F} = \big[\mathbf{e}_{a_1}, \ldots, \mathbf{e}_{a_{n_{\textrm{t}}}}\big]
=\mathbf{I}_{N_{\textrm{t}}}(:,\Omega_{\textrm{t}})
\in\{0,1\}^{N_{\textrm{t}}\times n_{\textrm{t}}}$, with $\mathbf{F}^{\mathsf H}\mathbf{F}=\mathbf{I}_{n_{\textrm{t}}}$ and where $\mathbf{e}_{a_i} \in \{0,1\}^{N_{\textrm{t}}}$ is the $a_i$-th canonical basis vector. The resulting transmitted pilot signal is $\mathbf{X}=\mathbf{F}\mathbf{S}\in\mathbb{C}^{N_{\textrm{t}}\times T}$.
The received signal is given by $\mathbf{Y}=\mathbf{H}\mathbf{X}+\mathbf{N}=\mathbf{H}\mathbf{F}\mathbf{S}+\mathbf{N}$, where $\mathbf{H}\in\mathbb{C}^{N_{\textrm{r}}\times N_{\textrm{t}}}$ denotes the channel matrix and $\mathbf{N}$ is the noise matrix with i.i.d. $\CN(0,\sigma_{\textrm{n}}^2)$ entries. De-spreading the pilots by right-multiplication with $\mathbf{S}^{\mathsf H}$ yields the observation matrix $\tilde{\mathbf{Y}} = \mathbf{Y}\mathbf{S}^{\mathsf H}=\mathbf{H}\mathbf{F}+\mathbf{N}\mathbf{S}^{\mathsf H}
\in \mathbb{C}^{N_{\textrm{r}}\times n_{\textrm{t}}}$.
Thus, $\tilde{\mathbf{Y}}$ contains noisy observations of the $n_{\textrm{t}}$ sounded columns of $\mathbf{H}$ indexed by $\Omega_{\textrm{t}}$~\cite{ShahNovel}, while the remaining $N_{\textrm{t}}-n_{\textrm{t}}$ channel columns remain unobserved and are inferred via the proposed SC-GPR method at the receiver, as detailed in the following section. Equivalently, these matrix observations correspond to noisy samples of individual channel entries indexed on the antenna grid, which we use as training set for GPR.

\section{Proposed GPR Framework With SC Kernel}
\label{sec:GPR}
This section introduces the proposed SC-GPR estimator, establishes its statistical properties, and presents a scalable inference routine for reduced-pilot CSI acquisition. Our approach adopts the proposed SC kernel for GPR, which: i) encodes the channel’s second-order structure; ii) removes hyperparameter training to reduce the computation cost; iii) yields a closed-form estimator with calibrated uncertainty; and iv) exploits model structure for scalability. The proposed SC kernel prior captures anisotropy and non-separable transmit-receive coupling implied by standard covariance models, which are not represented by conventional distance-based kernels. As a result, the proposed SC-GPR method enables more accurate channel reconstruction under sparse sounding.

\textit{\textbf{Training and test sets:}}
Define the antenna index grid $\mathcal{G} = \{(r,t): r\in\{1,\ldots,N_{\textrm{r}}\},\ t\in\{1,\ldots,N_{\textrm{t}}\}\}\subset\mathbb{N}^2$, which serves as the input domain of the GPR. Let the observed (training) index set be $\Ind=\{(r,a_i): r=1,\ldots,N_{\textrm{r}},\ i=1,\ldots,n_{\textrm{t}}\}\subset\mathcal{G}$, with cardinality $|\Ind|=P=N_{\textrm{r}}n_{\textrm{t}}$, while the unobserved (test) set is $\IndTest=\mathcal{G}\setminus\Ind$, with cardinality $|\IndTest|=M$. Each observed entry of $\tilde{\mathbf{Y}}$ corresponds to a noisy sample of a channel coefficient, i.e.,
$\tilde{Y}_{r,i}=H_{r,a_i}+w_{r,i}$ for $i=1,\ldots,n_{\textrm{t}}$,
with $w_{r,i}\sim\CN(0,\sigma_\textrm{n}^2)$. Thus, the training input locations are stacked as $\mathbf{Z}=[\mathbf{z}_1,\ldots,\mathbf{z}_P]^{\mathsf{T}}\in\mathbb{N}^{P\times 2}$, with $\mathbf{z}_n=(r_n,a_{i_n})\in\Ind$, and the corresponding complex-valued observations are collected in $\mathbf{h}=[h_1,\ldots,h_P]^{\mathsf{T}}\in\mathbb{C}^{P}$, with $h_n=\tilde{Y}_{r_n,i_n}$. We denote the GPR training set by $\mathcal{D} = (\mathbf{Z},\mathbf{h})$. Similarly, the test input locations are given by $\mathbf{Z}_\ast=[\mathbf{z}_{\ast,1},\ldots,\mathbf{z}_{\ast,M}]^{\mathsf{T}}\in\mathbb{N}^{M\times 2}$, with $\mathbf{z}_{\ast,m}=(r_{\ast,m},t_{\ast,m})\in\IndTest$, and the unknown test outputs are denoted by $\mathbf{h}_\ast=[h_{\ast,1},\ldots,h_{\ast,M}]^{\mathsf{T}}\in\mathbb{C}^{M}$, with $h_{\ast,m}=H_{r_{\ast,m},t_{\ast,m}}$. We model $\{H_{r,t}\}$ as a zero-mean proper complex Gaussian process (GP) over $\mathcal{G}$, with covariance specified by the proposed SC kernel described next.

\textit{\textbf{SC kernel definition:}}
Let $\mathbf{u}=\vect(\mathbf{H})\in\mathbb{C}^{N_{\textrm{r}}N_{\textrm{t}}}$ denote the column-wise vectorization of the channel matrix. We index the channel coefficient $H_{r,t}$ by $n = r + (t-1)N_{\textrm{r}}$, with $n\in\{1,\ldots,N_{\textrm{r}}N_{\textrm{t}}\}$ and $\mathbf{u}[n]=H_{r,t}$. Likewise, $(r',t')$ is mapped to $m = r' + (t'-1)N_{\textrm{r}}$, with $\mathbf{u}[m]=H_{r',t'}$. The channel covariance matrix is defined as $\mathbf{R}_{\textrm{H}} = \E\!\left[\mathbf{u}\mathbf{u}^{\mathsf{H}}\right]
\in \mathbb{C}^{N_{\textrm{r}}N_{\textrm{t}} \times N_{\textrm{r}}N_{\textrm{t}}}$,
whose $(n,m)$-th entry is given by $\mathbf{R}_{\textrm{H}}[n,m] = \E\!\left[H_{r,t} H_{r',t'}^{\ast}\right]\in \mathbb{C}$. We define the SC kernel as
\begin{equation}
k_{\textrm{SC}}\big((r,t),(r',t')\big) = \mathbf{R}_{\textrm{H}}[n,m],
\label{eq:kSC_def}
\end{equation}
which yields a Hermitian kernel suitable for proper complex GPR.
From $k_{\textrm{SC}}(\cdot)$, we construct the Gram matrices
$\mathbf{K}=k_{\textrm{SC}}(\mathbf{Z},\mathbf{Z})\in\mathbb{C}^{P\times P}$,
$\mathbf{K}_\ast=k_{\textrm{SC}}(\mathbf{Z},\mathbf{Z}_\ast)\in\mathbb{C}^{P\times M}$, and
$\mathbf{K}_{\ast\ast}=k_{\textrm{SC}}(\mathbf{Z}_\ast,\mathbf{Z}_\ast)\in\mathbb{C}^{M\times M}$~\cite{GPs_For_ML}.
The resulting GPR posterior for the test locations $\IndTest$ is
\begin{equation}
\widehat{\mathbf{h}}_\ast=\mathbf{K}_\ast^{\mathsf H}\big(\mathbf{K}+\sigma_{\textrm{n}}^2\mathbf{I}_P\big)^{-1}\mathbf{h} \in \mathbb{C}^{M\times 1},
\label{eq:GPR_post_mean}
\end{equation}
where $\mathbf{h}\in \mathbb{C}^{P\times 1}$ contains the complex-valued training observations, $\widehat{\mathbf{h}}_\ast\in \mathbb{C}^{M\times 1}$ stacks the posterior means, and $\boldsymbol{\Sigma}_\ast=\mathbf{K}_{\ast\ast}-\mathbf{K}_\ast^{\mathsf H}\big(\mathbf{K}+\sigma_{\textrm{n}}^2\mathbf{I}_P\big)^{-1}\mathbf{K}_\ast\in \mathbb{C}^{M\times M}$ is the posterior covariance~\cite{shah2025,GPs_For_ML}.

Next, Proposition~\ref{prop:sc-psd} proves that $k_{\textrm{SC}}(\cdot)$ is a valid kernel, i.e., for any finite index set, its Gram matrix is Hermitian PSD, ensuring well-posed GPR posteriors and numerically stable inference. Moreover, Proposition~\ref{prop:blup} shows that the SC-GPR posterior mean equals the LMMSE estimator induced by the same second-order model, establishing linear-optimality without assuming channel Gaussianity. In the following, we use the notations $\Cov(\mathbf{a})=\E[\mathbf{a}\mathbf{a}^{\mathsf H}]$ and $\Cov(\mathbf{a}, \mathbf{b})=\E[\mathbf{a}\mathbf{b}^{\mathsf H}]$.

\begin{proposition}[Positive semidefiniteness of $k_{\textrm{SC}}$]\label{prop:sc-psd}
Let $\mathbf{R}_{\textnormal{H}}$ be Hermitian PSD. For any finite set $\{\mathbf{z}_m\}_{m=1}^M\subset\mathcal{G}$, the Gram matrix with entries $[k_{\textrm{SC}}(\mathbf{z}_m,\mathbf{z}_{m'})]_{m,m'}$ is Hermitian PSD.
\end{proposition}
\begin{IEEEproof}
Index $\mathbf{z}_m=(r_m,t_m)$ by $i_m=(t_m-1)N_{\textrm{r}}+r_m$ and define $\mathbf{E}=[\mathbf{e}_{i_1},\ldots,\mathbf{e}_{i_M}]\in\mathbb{C}^{(N_{\textrm{r}}N_{\textrm{t}})\times M}$.
Then, we have $\mathbf{K}=\mathbf{E}^{\mathsf{H}}\mathbf{R}_{\textrm{H}}\mathbf{E}$, which is Hermitian. Furthermore, for any $\boldsymbol{\beta}\in\mathbb{C}^M$, we have $\boldsymbol{\beta}^{\mathsf H}\mathbf{K}\boldsymbol{\beta}
=(\mathbf{E}\boldsymbol{\beta})^{\mathsf{H}}\mathbf{R}_{\textrm{H}}(\mathbf{E}\boldsymbol{\beta})\ge 0$, since $\mathbf{R}_{\textrm{H}}$ is Hermitian PSD.
Hence, we obtain $\mathbf{K}\succeq\mathbf{0}$.
\end{IEEEproof}

\begin{proposition}[Posterior mean equals LMMSE]\label{prop:blup}
Let $\mathbf{u}$ be zero-mean with covariance $\mathbf{R}_{\textrm{H}}\succeq\mathbf{0}$. Define $\mathbf{y}=\mathbf{B}\mathbf{u}+\boldsymbol{\varepsilon}$, with $\E[\boldsymbol{\varepsilon}]=\mathbf{0}$, $\Cov(\boldsymbol{\varepsilon})=\sigma_\textrm{n}^2\mathbf{I}_P$, and $\E[\mathbf{u}\boldsymbol{\varepsilon}^{\mathsf H}]=\mathbf{0}$, where $\mathbf{B}\in\{0,1\}^{P\times N_{\textrm{r}}N_{\textrm{t}}}$ selects the $P$ observed entries. Among the linear estimators $\widehat{\mathbf{u}}=\mathbf{L}\mathbf{y}$, the LMMSE estimator that minimizes $\E\lVert\mathbf{u}-\widehat{\mathbf{u}}\rVert_2^2$ is
\begin{equation}
\widehat{\mathbf{u}}=\mathbf{R}_{\textrm{H}}\mathbf{B}^{\mathsf{H}}\big(\mathbf{B}\mathbf{R}_{\textrm{H}}\mathbf{B}^{\mathsf{H}}+\sigma_\textrm{n}^2\mathbf{I}_P\big)^{-1}\mathbf{y},
\label{eq:blup}
\end{equation}
with error covariance
$\mathbf{C}=\mathbf{R}_{\textrm{H}}-\mathbf{R}_{\textrm{H}}\mathbf{B}^{\mathsf{H}}(\mathbf{B}\mathbf{R}_{\textrm{H}}\mathbf{B}^{\mathsf{H}}+\sigma_\textrm{n}^2\mathbf{I}_P)^{-1}\mathbf{B}\mathbf{R}_{\textrm{H}}$.
A zero-mean proper complex GP prior whose kernel induces $\mathbf{R}_{\textrm{H}}$ (e.g., $k_{\textrm{SC}}(\cdot)$) yields a posterior mean that matches~\eqref{eq:blup}.
\end{proposition}

\begin{IEEEproof}
For $\widehat{\mathbf{u}}=\mathbf{L}\mathbf{y}$, the mean-square error is
$\E\|\mathbf{u}-\mathbf{L}\mathbf{y}\|_2^2
=\tr\big(\mathbf{R}_{\textrm{H}}-\Cov(\mathbf{u},\mathbf{y})\mathbf{L}^{\mathsf H}-\mathbf{L}\Cov(\mathbf{y},\mathbf{u})+\mathbf{L}\Cov(\mathbf{y})\mathbf{L}^{\mathsf H}\big)$,
with $\Cov(\mathbf{y},\mathbf{u})=\mathbf{B}\mathbf{R}_{\textrm{H}}$, $\Cov(\mathbf{u},\mathbf{y})=\mathbf{R}_{\textrm{H}}\mathbf{B}^{\mathsf H}$, and
$\Cov(\mathbf{y})=\mathbf{B}\mathbf{R}_{\textrm{H}}\mathbf{B}^{\mathsf{H}}+\sigma_\textrm{n}^2\mathbf{I}_P$.
Differentiating with respect to $\mathbf{L}$ and setting the gradient to zero gives
$\mathbf{L}^\star=\mathbf{R}_{\textrm{H}}\mathbf{B}^{\mathsf{H}}(\mathbf{B}\mathbf{R}_{\textrm{H}}\mathbf{B}^{\mathsf{H}}+\sigma_\textrm{n}^2\mathbf{I}_P)^{-1}$,
which yields~\eqref{eq:blup}. With a GP prior inducing $\mathbf{R}_{\textrm{H}}$, the GP posterior mean has the same form, so $k_{\textrm{SC}}(\cdot)$ coincides with LMMSE.
\end{IEEEproof}
Comparing~\eqref{eq:GPR_post_mean} with~\eqref{eq:blup} shows that $k_{\textrm{SC}}(\cdot)$ implements the covariance-induced LMMSE estimator under the reduced observation model. Under jointly proper complex Gaussian assumptions, it also coincides with the MMSE estimator. In addition, $k_{\textrm{SC}}(\cdot)$ provides calibrated per-entry and joint uncertainty via $\boldsymbol{\Sigma}_\ast$.

\textit{\textbf{Complex channel reconstruction:}}
The complex GP posterior in~\eqref{eq:GPR_post_mean} directly provides $\widehat{H}_{r,t}$ for each $(r,t)\in\IndTest$ together with its posterior uncertainty.

\begin{table}[t]
\centering
\renewcommand{\arraystretch}{1.15}
\setlength{\tabcolsep}{4pt}
\begin{tabular}{|p{1.5cm}|p{1.45cm}|p{5.0cm}|}
\hline
\textbf{Estimator} & \textbf{Complexity} & \textbf{Remarks} \\
\hline
\textbf{SC-GPR (proposed)}
& $\mathcal{O}(P^{3})$
& $\mathbf{K}$ is indexed from $\mathbf{R}_{\textrm{H}}$ in~\eqref{eq:kSC_def}; no hyperparameter loop; one Cholesky solve of $\mathbf{K}+\sigma_{\textrm{n}}^2\mathbf{I}_P$. \\
\hline
\textbf{Learning-based GPR~\cite{shah2025LOWoverhead, ShahNovel}} 
& $\mathcal{O}(QP^{3})$
& $Q$ marginal-likelihood iterations; each builds the $P\times P$ kernel, factors $\mathbf{K}+\sigma_{\textrm{n}}^2\mathbf{I}_P$, and computes gradients. \\
\hline
\textbf{LMMSE}
& $\mathcal{O}\big((N_{\textrm{r}}N_{\textrm{t}})^{3}\big)$
& Solves a system of size $N_{\textrm{r}}N_{\textrm{t}}$; no calibrated uncertainty. \\
\hline
\textbf{LS}
& $\mathcal{O}(N_{\textrm{r}}N_{\textrm{t}}T)$
& Point estimates only; lowest cost but noise-limited; no spatial prior or uncertainty. \\
\hline
\end{tabular}
\caption{Computational complexity comparison.} \vspace{-1mm}
\label{tab:complexity}
\end{table}

\textit{\textbf{Computational complexity:}}
SC-GPR improves efficiency compared with learning-based GPR~\cite{Jieao2025IEEE_TIT, shah2025LOWoverhead, ShahNovel} and the LS and LMMSE baselines with full-array training. Specifically, since $P=N_{\textrm{r}}n_{\textrm{t}}$ is the number of observed channel entries, computing the SC-GPR posterior in~\eqref{eq:GPR_post_mean} requires forming the corresponding $P\times P$ Gram matrix and solving a single linear system via Cholesky factorization, resulting in a computational complexity of $\mathcal{O}(P^{3})$. This cost arises from the inversion (or factorization) of $\mathbf{K}+\sigma_{\textrm{n}}^2\mathbf{I}_P$ and dominates all the other operations. In learning-based GPR~\cite{Jieao2025IEEE_TIT, shah2025LOWoverhead, ShahNovel}, the hyperparameters are learned via marginal-likelihood maximization. Each of the $Q$ outer iterations assembles the $P\times P$ kernel matrix, factors $\mathbf{K}+\sigma_{\textrm{n}}^2\mathbf{I}_P$, and evaluates the kernel gradients, leading to an overall complexity of $\mathcal{O}(Q P^{3})$, as summarized in Table~\ref{tab:complexity}. In contrast, the kernel $k_{\textrm{SC}}(\cdot)$ in SC-GPR is fully determined by the channel covariance $\mathbf{R}_{\textrm{H}}$, the Gram matrices are obtained by direct indexing, and no outer optimization loop is required. Compared with the LMMSE estimator with full-array training, which involves solving a linear system of dimension $N_{\textrm{r}}N_{\textrm{t}}$ and thus scales as $\mathcal{O}\big((N_{\textrm{r}}N_{\textrm{t}})^{3}\big)$, SC-GPR achieves a strictly lower computational complexity for $n_{\textrm{t}} < N_{\textrm{t}}$. Hence, SC-GPR provides: i) an LMMSE-equivalent posterior mean implied by $\mathbf{R}_{\textrm{H}}$; ii) calibrated per-entry uncertainty from the GP posterior; and iii) scalable computation that explicitly exploits the channel structure.

\section{Numerical Evaluation}
We evaluate the proposed SC-GPR framework under reduced-pilot training to quantify its performance. In particular, we consider three pilot subsampling factors, i.e., $\Delta \in \{2,3,4\}$, corresponding to $50\%$, $67\%$, and $75\%$ reduced training overheads, respectively. Unless stated otherwise, the simulation parameters are summarized in Table~\ref{tab:params}.

We compare the proposed SC-GPR method with the learning-based GPR estimators from~\cite{shah2025LOWoverhead} employing distance-based kernels, namely the radial basis function (RBF)
$
k_{\textrm{RBF}}(\mathbf{z},\mathbf{z}_\ast)=\gamma_{\textrm{RBF}}\exp \bigl(-\|\mathbf{z}-\mathbf{z}_\ast\|_{2}^2/(2\ell_{\textrm{RBF}}^2)\bigr)
$,
the Mat\'ern
$
k_{\textrm{Mat}}(\mathbf{z},\mathbf{z}_\ast)=\gamma_{\textrm{Mat}}\frac{2^{1-\nu}}{\Gamma(\nu)}
\bigl(\tfrac{\sqrt{2\nu}\|\mathbf{z}-\mathbf{z}_\ast\|_{2}}{\ell_{\textrm{Mat}}}\bigr)^{\nu}
K_\nu\bigl(\tfrac{\sqrt{2\nu}\|\mathbf{z}-\mathbf{z}_\ast\|_{2}}{\ell_{\textrm{Mat}}}\bigr)
$,
and the rational quadratic (RQ)
$
k_{\textrm{RQ}}(\mathbf{z},\mathbf{z}_\ast)=\gamma_{\textrm{RQ}}
\bigl(1+\|\mathbf{z}-\mathbf{z}_\ast\|_{2}^2/(2\alpha \ell_{\textrm{RQ}}^2)\bigr)^{-\alpha}
$.
Their hyperparameters $\gamma_{\textrm{RBF}},\gamma_{\textrm{Mat}},\gamma_{\textrm{RQ}}$ and
$\ell_{\textrm{RBF}},\ell_{\textrm{Mat}},\ell_{\textrm{RQ}}$ denote the kernel scales and lengthscales, respectively, while $\nu$ and $\alpha$ control the Mat\'ern smoothness and the RQ multiscale behavior, respectively. All the relevant hyperparameters are learned via marginal-likelihood maximization as outlined in~\cite{shah2025LOWoverhead}. In the comparison, we also include the LS and LMMSE estimators with full-array orthogonal training, which serve as the conventional (non-GPR) performance benchmarks.

\begin{table}[t]
\centering
\renewcommand{\arraystretch}{1.1}
\resizebox{\columnwidth}{!}{%
\begin{tabular}{|l|c|c|}
\hline
\textbf{System Parameter} & \textbf{Symbol} & \textbf{Value} \\
\hline
\multicolumn{3}{|c|}{\textbf{Array and propagation}} \\
\hline
Receive $\times$ transmit antennas & $N_{\textrm{r}}\times N_{\textrm{t}}$ & $36\times36$ \\
Antenna activation stride & $\Delta$ & $\{2, 3, 4\}$ \\
Element spacing relative to wavelength &  & $0.5$ \\ 
Carrier frequency & & $\SI{28}{GHz}$ \\ 
Azimuth angle spread &  & $\pi/6\, \mathrm{rad}$ \\ 
\hline
\multicolumn{3}{|c|}{\textbf{Training / SNR model}} \\
\hline
Pilot lengths for SC-GPR & $T=n_{\textrm{t}}$ & $\{18,12,9\}$ \\
Pilot length for the baselines & $T=N_{\textrm{t}}$ & $36$ \\
SNR & $\rho$ & $0$~dB \\ 
\hline
\multicolumn{3}{|c|}{\textbf{Learning-based GPR hyperparameters~\cite{shah2025LOWoverhead}}} \\
\hline
Kernel scale (init; bounds) & $\gamma_{\textrm{RBF}},\gamma_{\textrm{Mat}},\gamma_{\textrm{RQ}}$ & $1.0$; $[10^{-2},10^{2}]$ \\
RBF lengthscale (init; bounds) & $\ell_{\textrm{RBF}}$ & $0.5$; $[10^{-2},10]$ \\
Mat\'ern lengthscale (init; bounds) & $\ell_{\textrm{Mat}}$ & $0.5$; $[10^{-2},10]$ \\
Mat\'ern smoothness & $\nu$ & $1.5$ (fixed) \\
RQ lengthscale (init; bounds) & $\ell_{\textrm{RQ}}$ & $0.5$; $[10^{-2},5]$ \\
RQ shape (init; bounds) & $\alpha$ & $0.5$; $[10^{-1},5]$ \\
\hline
\end{tabular}%
}
\caption{Simulation setup and hyperparameters.} \vspace{-1mm}
\label{tab:params}
\end{table}

To evaluate the robustness of the proposed SC-GPR with respect to spatial channel statistics, we consider two channel models for $\mathbf{R}_{\textrm{H}}$: the separable Kronecker model~\cite{kronecker2002} and the non-separable Weichselberger model~\cite{weichselberger2006}. Under the Kronecker model, the channel covariance is separable across the transmit- and receive-side, assuming independent spatial correlations at the transmitter and receiver~\cite{kronecker2002}. In contrast, the Weichselberger model yields a generally non-separable covariance matrix that captures joint transmit-receive coupling through a full power-coupling structure in a common eigenbasis, representing a more general and challenging propagation environment. To assess estimation accuracy, uncertainty calibration, and system-level communication performance, we employ three evaluation metrics, described next.

\subsection{Prediction Error and Empirical $95\%$ Error Ellipses}
\begin{figure*}[tb]
  \centering
  \begin{subfigure}{0.49\textwidth}
    \includegraphics[width=\linewidth,trim=0.01pt 0 0.01pt 0,clip]{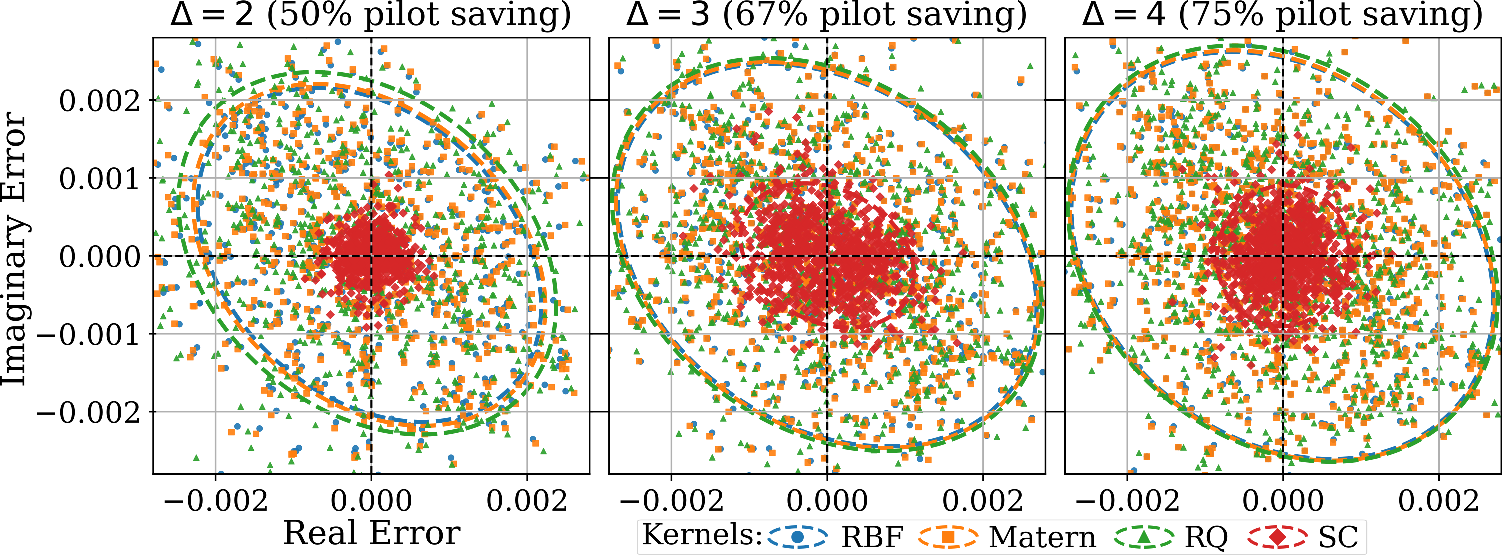}
    \caption{Separable covariance structure from the Kronecker model.}
    \label{fig:kron}
  \end{subfigure}\hfill
  \begin{subfigure}{0.49\textwidth}
    \includegraphics[width=\linewidth,trim=0.01pt 0 0.01pt 0,clip]{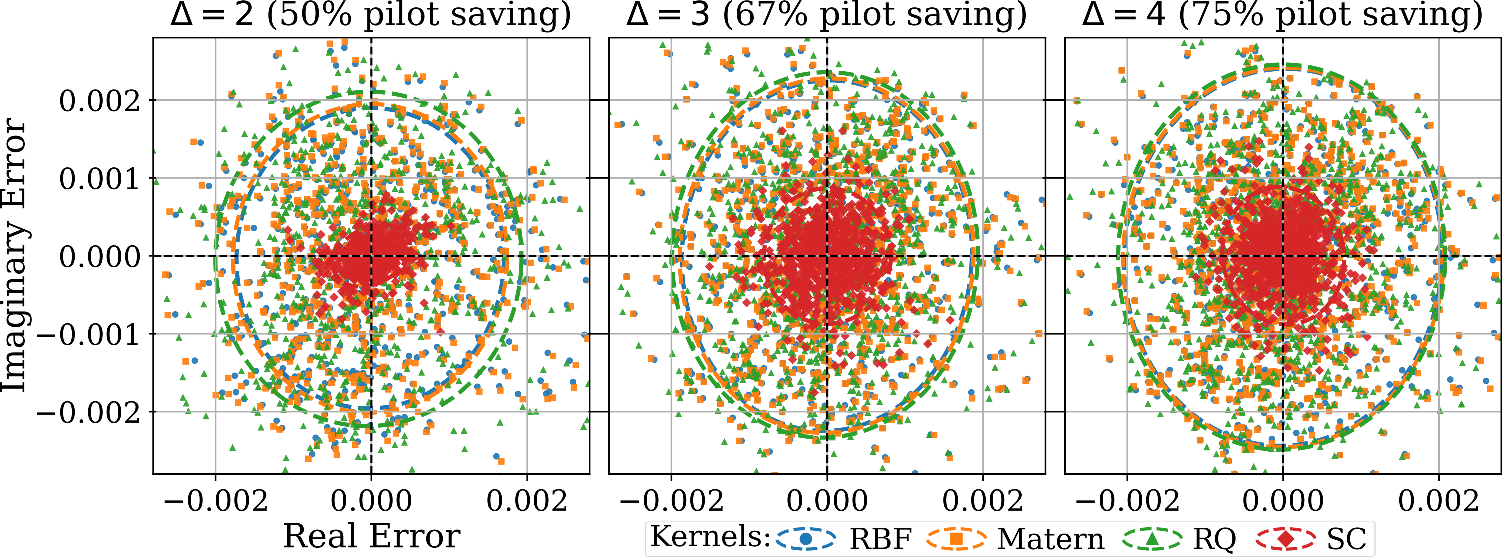}
    \caption{Joint-eigenbasis covariance structure from the Weichselberger model.}
    \label{fig:weich}
  \end{subfigure}
  \caption{Each scatter plot shows $\Real(\epsilon_{ij})$ versus $\Imag(\epsilon_{ij})$ per complex channel entry with the empirical $95\%$ error ellipse; smaller, more isotropic ellipses indicate lower error variance and weaker residual correlation.} \vspace{-1mm}
  \label{fig:ScatterError}
\end{figure*}
We first evaluate the entry-wise complex prediction error
$\epsilon_{ij} = H_{ij} - \widehat{H}_{ij}$ in Fig.~\ref{fig:ScatterError}, which is visualized via scatter plots of $\Real(\epsilon_{ij})$ versus $\Imag(\epsilon_{ij})$ together with the empirical $95\%$ error ellipses.
We compare the proposed SC-GPR method against the RBF, Mat\'ern, and RQ kernel-based GPR baselines from~\cite{shah2025LOWoverhead} under both the separable Kronecker~\cite{kronecker2002} and the non-separable Weichselberger~\cite{weichselberger2006} channel models, as shown in Figs.~\ref{fig:kron} and~\ref{fig:weich}, respectively. Tighter clustering of the scatter points around the origin indicates lower estimation error, whereas wider dispersion corresponds to higher error variance.
Across all pilot subsampling factors $\Delta \in \{2,3,4\}$, corresponding to $\{50\%,67\%,75\%\}$ pilot savings, it is evident that the proposed SC-GPR consistently achieves the smallest error dispersion among all considered methods. This performance gain stems from SC-GPR explicitly encoding the underlying spatial propagation statistics through $\mathbf{R}_{\textrm{H}}$, which generic distance-based kernels are unable to capture accurately, particularly under severe pilot scarcity. The overlaid $95\%$ error ellipses summarize the empirical bivariate distribution of the pooled complex estimation errors.
The ellipse area reflects the total error variance, while its orientation and eccentricity capture correlation and anisotropy between the real and imaginary components; larger ellipses therefore indicate higher uncertainty and lower estimation precision. For all reduced-pilot configurations, SC-GPR yields the smallest empirical ellipse areas, confirming superior uncertainty concentration. Moreover, ellipses under the Weichselberger model are generally more isotropic than those under the Kronecker model, owing to the joint-eigenbasis structure of the Weichselberger covariance, which mitigates residual real-imaginary coupling.

\subsection{Uncertainty Calibration ($95\%$ coverage)}
We assess uncertainty calibration via the empirical coverage of marginal $95\%$ credible intervals for each channel coefficient. For each entry, we compute
$\Real(\widehat{H}_{ij}) \pm 1.96\,\sigma^{(\Real)}_{ij}$ and
$\Imag(\widehat{H}_{ij}) \pm 1.96\,\sigma^{(\Imag)}_{ij}$,
where $\sigma^{(\Real)}_{ij}$ and $\sigma^{(\Imag)}_{ij}$ are the posterior standard deviations. The empirical coverage is the fraction of true values within these intervals, averaged over the real and imaginary components. For brevity, we compare against the competitive RQ kernel. Table~\ref{tab:ellipse_compact} reports the ellipse area $\mathrm{Area}_{95}$, axis ratio, and empirical coverage probability. The axis ratio, defined as the ratio between the major and minor semi-axes of the empirical $95\%$ error ellipse, is computed from the eigenvalues of the $2\times2$ covariance matrix of $(\Real(\epsilon_{ij}),\Imag(\epsilon_{ij}))$. Values closer to unity indicate more isotropic and weakly correlated uncertainty. SC-GPR improves coverage relative to baselines and remains reasonably calibrated across all pilot-saving regimes, indicating reliable calibration under reduced training overhead. By contrast, the RQ baseline exhibits similar coverage due to learned lengthscales becoming large relative to the array aperture under sparse sounding, resulting in near-constant correlations and comparable posterior variances. Across both separable and non-separable models, SC-GPR achieves tighter uncertainty concentration and comparable or improved isotropy. These results confirm that SC-GPR provides well-calibrated uncertainty estimates in addition to accurate point estimation. Owing to its greater physical realism, subsequent analysis focuses on the Weichselberger model~\cite{weichselberger2006}.

\begin{table}[t]
\centering
\setlength{\tabcolsep}{2.2pt}
\renewcommand{\arraystretch}{1.05}
\resizebox{\columnwidth}{!}{%
\begin{tabular}{|l|l|r|r|r|r|r|r|}
\hline
\multirow{2}{*}{\textbf{Model}} & \multirow{2}{*}{\textbf{$\Delta$}}
& \multicolumn{2}{|c|}{\textbf{Error variance ({Area}$_{95})$}} 
& \multicolumn{2}{c|}{\textbf{Axis ratio}} 
& \multicolumn{2}{c|}{\textbf{Coverage}} \\
\cline{3-8}
& & \textbf{RQ} & \textbf{SC} & \textbf{RQ} & \textbf{SC} & \textbf{RQ} & \textbf{SC} \\
\hline
\multirow{3}{*}{Kronecker}
& $2$
& $1.70e-5$ & $5.93e-7$
& $1.348$ & $1.101$ & $0.888$ & $0.898$ \\
& $3$
& $2.10e-5$ & $3.13e-6$
& $1.345$ & $1.376$ & $0.906$ & $0.909$ \\
& $4$
& $2.25e-5$ & $2.41e-6$
& $1.261$ & $1.075$ & $0.923$ & $0.907$ \\
\hline
\multirow{3}{*}{Weichselberger}
& $2$
& $1.32e-5$ & $5.86e-7$
& $1.093$ & $1.454$ & $0.886$ & $0.899$ \\
& $3$
& $1.44e-5$ & $2.23e-6$
& $1.196$ & $1.069$ & $0.917$ & $0.915$ \\
& $4$
& $1.63e-5$ & $2.16e-6$
& $1.178$ & $1.140$ & $0.928$ & $0.907$ \\
\hline
\end{tabular}}
\caption{Empirical $95\%$ error-ellipse metrics comparing baseline RQ and the proposed SC-GPR estimator (denoted by SC).} \vspace{-1mm}
\label{tab:ellipse_compact}
\end{table}

\subsection{Spectral Efficiency}

\begin{figure}[t]
    \centering
    \includegraphics[width=0.85\linewidth]{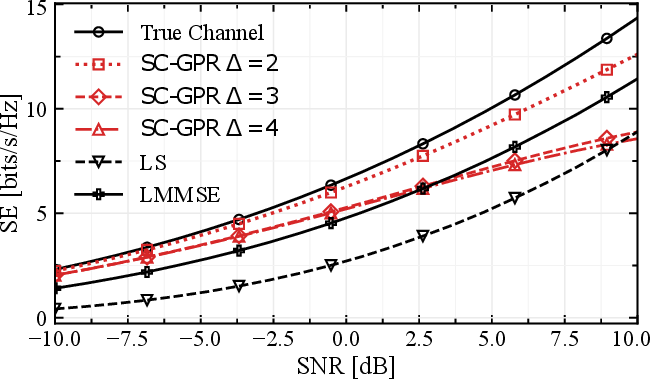}
    \caption{SE versus SNR comparing the proposed SC-GPR estimator with reduced overheads against ground-truth and LS/LMMSE baselines.} \vspace{-1mm}
    \label{fig:SE_SC-GPR}
\end{figure}

\begin{figure}[t]
    \centering
    \includegraphics[width=0.85\linewidth]{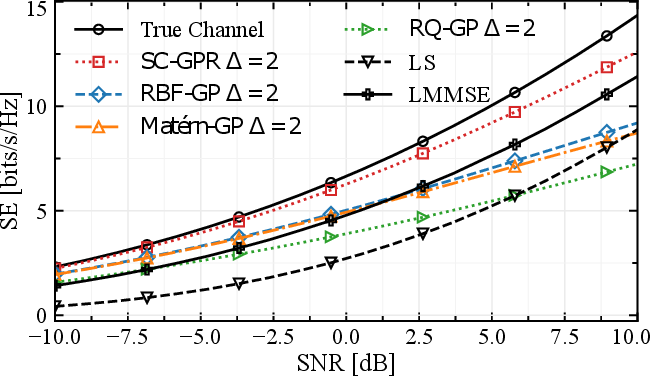}
    \caption{SE versus SNR comparing the proposed SC-GPR estimator with distance-based kernels under $50\%$ pilot savings and LS/LMMSE under full-array training.} \vspace{-1mm}
    \label{fig:SE_all_kernels}
\end{figure}

To quantify the system-level impact, we evaluate the SE of a multi-stream MIMO link using a linear receiver designed from the estimated channel. For a channel estimate $\widehat{\mathbf{H}}\in\mathbb{C}^{N_{\textrm{r}}\times N_{\textrm{t}}}$, we adopt the LMMSE detector
$\mathbf{W}(\widehat{\mathbf{H}})= \big( \widehat{\mathbf{H}}\widehat{\mathbf{H}}^{\mathsf{H}} + \frac{N_{\textrm{t}}}{\rho}\mathbf{I}_{N_{\textrm{r}}} \big)^{-1}\widehat{\mathbf{H}}=[\mathbf{w}_1,\ldots,\mathbf{w}_{N_{\textrm{t}}}]$.
Let $\mathbf{h}_{k}$ and $\mathbf{w}_{k}$ denote the $k$th columns of the true channel $\mathbf{H}$ and detector $\mathbf{W}(\widehat{\mathbf{H}})$, respectively. The post-equalization signal-to-interference-plus-noise ratio (SINR) of stream $k$ is
\begin{equation}
\mathrm{SINR}_{k}(\widehat{\mathbf{H}})
= \frac{\bigl|\mathbf{w}_{k}^{\mathsf{H}}\mathbf{h}_{k}\bigr|^2}
{\sum_{j\neq k}\bigl|\mathbf{w}_{k}^{\mathsf{H}}\mathbf{h}_j\bigr|^2 
+ \frac{N_{\textrm{t}}}{\rho}\|\mathbf{w}_{k}\|_{2}^2},
\end{equation}
where $\widehat{\mathbf{H}}$ affects the performance only through the detector $\mathbf{W}(\widehat{\mathbf{H}})$. The SE is defined as
$\mathrm{SE}(\widehat{\mathbf{H}})
= (1-T/T_{\textrm{c}})\E\big[\sum_{k=1}^{N_{\textrm{t}}} \log_2\bigl(1+\mathrm{SINR}_{k}(\widehat{\mathbf{H}})\bigr)\big]$, where $T_{\textrm{c}}$ is the coherence block length and $T$ is the pilot length.
We evaluate $\mathrm{SE}(\cdot)$ using the true channel $\mathbf{H}$, the GPR estimate $\widehat{\mathbf{H}}_{\textrm{GPR}}$, and the LS/LMMSE baselines $\widehat{\mathbf{H}}_{\textrm{LS}}$ and $\widehat{\mathbf{H}}_{\textrm{LMMSE}}$. For LS and LMMSE, we use full-array orthogonal training with $T \ge N_{\textrm{t}}$ and pilot matrix $\mathbf{S} \in \mathbb{C}^{N_{\textrm{t}}\times T}$. The LS estimate is given by
$
\widehat{\mathbf{H}}_{\textrm{LS}} = \mathbf{Y}\mathbf{S}^{\mathsf{H}} (\mathbf{S}\mathbf{S}^{\mathsf{H}})^{-1}.
$
Defining $\mathbf{y}= \vect(\mathbf{Y})$, the LMMSE estimate is given by
$
\widehat{\mathbf{u}}_{\textrm{LMMSE}} = \mathbf{R}_{\textrm{H}}\mathbf{A}^{\mathsf{H}}(\mathbf{A}\mathbf{R}_{\textrm{H}}\mathbf{A}^{\mathsf{H}}+\sigma_\textrm{n}^2\mathbf{I}_{T N_{\textrm{r}}})^{-1}\mathbf{y}
$
with $\mathbf{A}= \mathbf{S}^{\mathsf{T}}\otimes \mathbf{I}_{N_{\textrm{r}}}$ and $\widehat{\mathbf{H}}_{\textrm{LMMSE}}= \mathrm{unvec}(\widehat{\mathbf{u}}_{\textrm{LMMSE}})$, where $\mathrm{unvec}$ is the inverse vectorization operator. Fig.~\ref{fig:SE_SC-GPR} reports the SE achieved by the proposed SC-GPR estimator under pilot subsampling factors
$\Delta\in\{2,3,4\}$, corresponding to $\{50\%,67\%,75\%\}$ pilot savings.
Even with $75\%$ pilot savings, SC-GPR consistently outperforms LS across the SNR range.
Moreover, for low SNR, SC-GPR with $\Delta\in\{3,4\}$ achieves higher SE than LMMSE with full-array training; for higher SNR, the latter improves and eventually surpasses the reduced-pilot schemes.
With $50\%$ pilot savings, SC-GPR closely tracks the SE with the true channel and substantially outperforms LMMSE with full-array training.
This highlights that exploiting the spatial covariance structure via the SC kernel can yield superior system-level performance even with reduced training overhead. Fig.~\ref{fig:SE_all_kernels} compares SC-GPR against learning-based GPR with standard distance-based kernels under $50\%$ pilot savings,
while LS and LMMSE with $0\%$ pilot savings are included as baselines. SC-GPR most closely matches the SE with the true channel. Among the generic kernels, RBF performs best, followed by Mat\'ern and RQ.

\subsection{Discussion}
Table~\ref{tab:pilot_MI_GPR} summarizes the trade-off between pilot overhead and performance. The proposed SC-GPR estimator improves estimation fidelity, preserves link-level performance, and reduces training overhead by exploiting $\mathbf{R}_{\textrm{H}}$ through the SC kernel. It avoids marginal-likelihood hyperparameter optimization and the associated $Q$-iteration cost, it scales as $\mathcal{O}(P^3)$ with $P=N_{\textrm{r}}n_{\textrm{t}}$ rather than $N_{\textrm{r}}N_{\textrm{t}}$, and it degrades gracefully as pilot savings increase, whereas learning-based distance kernels are more sensitive to sparse sounding~\cite{shah2025LOWoverhead}. Consequently, SC-GPR consistently outperforms LS and approaches LMMSE with full-array training despite the substantially lower training overhead and computational burden.

\begin{table}[tb]
\centering
\renewcommand{\arraystretch}{1.1}
\setlength{\tabcolsep}{2.2pt}
\begingroup\color{black}
\begin{tabular}{|l|c|c|c|c|c|}
\hline
\textbf{Estimator} 
& \textbf{$\Delta$} 
& \textbf{$T$ / saving} 
& \textbf{Relative SE [\%]} 
& \textbf{NMSE [dB]} 
& \textbf{Complexity} \\
\hline
SC-GPR 
& $2$ 
& $18$ / \textbf{$50\%$} 
& $94.5$ 
& $-10.45$ 
& $\mathcal{O}(648^{3})$ \\
SC-GPR 
& $3$ 
& $12$ / $67\%$ 
& $80.0$
& $-9.12$ 
& $\mathcal{O}(432^{3})$ \\
SC-GPR 
& $4$ 
& $9$ / $75\%$ 
& $79.0$ 
& $-8.35$ 
& $\mathcal{O}(324^{3})$ \\
\hline
RBF-GPR 
& $2$ 
& $18$ / \textbf{$50\%$} 
& $76.1$ 
& $-2.81$ 
& $\mathcal{O}(Q \times 648^{3})$ \\
RBF-GPR 
& $3$ 
& $12$ / $67\%$ 
& $61.3$
& $-1.67$ 
& $\mathcal{O}(Q \times 432^{3})$ \\
RBF-GPR 
& $4$ 
& $9$ / $75\%$ 
& $46.3$ 
& $-1.18$ 
& $\mathcal{O}(Q \times 324^{3})$ \\
\hline
LS 
& -- 
& $36$ / $0\%$ 
& $49.5$ 
& $-4.96$ 
& $\mathcal{O}(36^{3})$ \\
LMMSE 
& -- 
& $36$ / $0\%$ 
& $73.90$ 
& $-10.49$
& $\mathcal{O}(1296^{3})$ \\
\hline
\end{tabular}
\endgroup
\caption{Comparison of the proposed SC-GPR estimator with the RBF-GPR, LS, and LMMSE estimators in terms of pilot overhead, relative SE, NMSE, and computational complexity.} \vspace{-1mm}
\label{tab:pilot_MI_GPR}
\end{table}

\section{Conclusions}
We proposed a GPR-based channel estimation framework utilizing the SC kernel constructed directly from the second-order channel statistics. The resulting estimator admits a closed-form posterior mean and covariance, eliminates marginal-likelihood hyperparameter learning, and enables uncertainty-aware inference via credible regions. We established that the SC kernel is PSD and showed that the resulting posterior mean coincides with the corresponding LMMSE estimator under the same second-order statistics, without requiring Gaussian channel distributions.
Numerical results under both separable and non-separable correlation models demonstrated that SC-GPR achieves substantially lower NMSE
and better preservation of SE under aggressive pilot reduction, while maintaining lower computational cost than learning-based GPR baselines and LMMSE with full-array training.
Future work will address robustness under covariance mismatch and the extension to time-varying channels.
\bibliographystyle{IEEEtraN_renamed}
\bibliography{bibliography}
\end{document}